\documentclass[10pt]{article}
\usepackage{amsfonts}
\usepackage{amssymb}

\newcommand{\C}{{\bf C}}
\newcommand{\text}{\mbox}
\newcommand{\oc}{{\sf{oc}}}

\renewcommand{\O}{{\bf O}}
\renewcommand{\H}{{\bf H}}
\newcounter{thaler}
\newenvironment{mlist}{\begin{list}{\arabic{thaler}}%
{\usecounter{thaler}
\setlength{\rightmargin}{\leftmargin}
\topsep=0pt
\itemsep=0pt
\parskip=0pt
\parsep=0pt
}}{\end{list}}

\begin{document}

\begin{center}{\large \sc Topological Test Spaces}\footnote{I wish to 
dedicate this paper to the memory of Frank J. Hague III}\\
Alexander Wilce\\ Department of Mathematical Sciences,  
Susquehanna University \\ Selinsgrove, Pa 17870  \ email: wilce@susqu.edu \end{center}

\begin{abstract}  
A {\em test space} is the set of outcome-sets associated with a collection of experiments. 
This notion provides a simple mathematical framework for the study of probabilistic theories 
-- notably, quantum mechanics -- in which one is faced with incommensurable random quantities. 
In the case of quantum mechanics, the relevant test space, the set of orthonormal bases of a Hilbert space, carries 
significant topological structure. This paper inaugurates a general study of topological 
test spaces. Among other things, we show that any topological test space with a compact space of outcomes is 
of finite rank. We also generalize results of Meyer and Clifton-Kent by showing that, 
under very weak assumptions, any second-countable topological 
test space contains a dense semi-classical test space. \end{abstract} 
                 
{\large \sc 0. Introduction}

A {\em test space} in the sense of Foulis and Randall [3, 4, 5], is a pair 
$(X,{\frak A})$ where $X$ is a non-empty set and ${\frak A}$ is a covering of $X$  
by non-empty subsets. \footnote{It is also usual to 
assume that ${\frak A}$ is {\em irredundant}, i.e., that no set in
${\frak A}$ properly contain another. For convenience, 
we relax this assumption.} 
The intended interpretation is that 
each set $E \in {\frak A}$ represents an exhaustive set of mutually exclusive 
possible {\em outcomes},  as of some experiment, decision, physical 
process, or {\em test}.  A {\em state}, or {\em probability weight}, on $(X,{\frak A})$  is a 
mapping $\omega : X \rightarrow [0,1]$ summing to $1$ over each test. 

Obviously, this framework subsumes discrete classical probability theory,  
which deals with test spaces $(E,\{E\})$ having only a single test. 
It also accommodates quantum probability theory, as follows. Let $\H$ be a Hilbert space, 
let $S = S(\H)$ be the unit sphere of $\H$, and let ${\frak F} = {\frak  F}(\H)$ 
denote the collection of all {\em frames}, i.e., maximal pairwise orthogonal subsets of $S$. 
The test space $(S,{\frak F})$ is a model for the set of maximally informative, discrete  quantum-mechanical 
experiments. As long as $\dim(\H) > 2$, Gleason's theorem  [6] 
tells us that every state $\omega$ on $(S,{\frak F})$ arises from a density 
operator $W$ on $\H$ via the rule $\omega(x) = \langle  Wx,x\rangle$ for 
all $x \in S$. 
                                
In this last example, the test space has a natural  topological structure: 
$S$ is a metric space, and ${\frak F}$ can be  topologized as well 
(in several ways).  The purpose of this paper is to provide a framework for 
the study of topological test spaces generally. Section 1 develops basic 
properties of the Vietoris topology, which we use heavily in the sequel. 
Section 2 considers topological test spaces in general, and locally finite 
topological test spaces in particular.  Section 3 addresses the problem of topologizing the logic of an 
algebraic topological test space. In section 4, we generalize results 
of Meyer [8] and Clifton and Kent [2] by showing that any second-countable topological 
test space satisfying a rather natural condition contains a dense semi-classical subspace. 
The balance of this section collects some essential background information concerning test  spaces (see [11] for a detailed 
survey). Readers familiar with this material can proceed directly to section 1.

{\bf 0.1 Events} Let $(X,{\frak A})$ be a test space. Two outcomes $x, y \in 
X$ are said  to be {\em orthogonal}, or {\em mutually exclusive}, if they are 
distinct  and belong to a common test. In this case, we write $x \perp y$. 
More  generally, a set $A \subseteq X$ is called an {\em event} for $X$ if 
there  exists a test $E \supseteq A$. The set of events is denoted  by ${\cal 
E}(X,{\frak A})$. 

There is a natural orthogonality relation on ${\cal E}(X,{\frak A})$ 
extending that on $X$, namely, $A \perp B$ iff $A \cap B = \emptyset$ and $A 
\cup B \in {\cal  E}(X,{\frak A})$. Every state $\omega$ on $(X,{\frak A})$ extends to a 
mapping $\omega : {\cal E}(X,{\frak A}) \rightarrow [0,1]$  given by 
$\omega(A) = \sum_{x \in A} \omega(x)$. If  $A \perp B$, then $\omega(A \cup B) 
= \omega(A) + \omega(B)$ for every probability weight $\omega$. 
Two events $A$ and $C$  are {\em complementary} -- abbreviated 
$A \oc C$ -- if they partition a test, and {\em perspective} if they are complementary to a common third event $C$. 
In this case, we write $A \sim B$. Note that if $A$ and $B$ are perspective, 
then for every state $\omega$ on $(X,{\frak A})$, $\omega(A) = 1 - \omega(C) = \omega(B)$.

{\bf 0.2 Algebraic Test Spaces} We say that $X$ is {\em algebraic} iff 
for all events $A, B, C \in {\cal E}(X,{\frak A})$, $A \sim B \ \text{and} \ B \oc C \ \Rightarrow \ A \oc C$. In this case, $\sim$ is an equivalence relation on ${\cal E}(X)$. Moreover, if $A \perp B$ and $B \sim C$, then $A \perp C$ as well, and $A \cup B \sim A 
\cup C$. 

Let $\Pi(X,{\frak A}) = {\cal E}(X,{\frak A})/\sim$, and write $p(A)$ for  
the $\sim$-equivalence class of an event $A \in {\cal E}(X)$. Then 
$\Pi$ carries a well-defined orthogonality relation, namely $p(A) \perp  
p(B) \Leftrightarrow A \perp B$, and also a partial binary operation $p(A) 
\oplus p(B) = p(A \cup B)$, defined for orthogonal pairs. We may also define 
$0 := p(\emptyset)$, $1 := p(E)$, $E \in {\frak A}$, and $p(A)' = p(C)$ where 
$C$ is any event complementary to $A$. 

The structure $(\Pi,\oplus, ', 0, 1)$, called the {\bf logic} of $(X,{\frak A})$, satisfies the following conditions: 
\begin{mlist} \item[(1)] $p \oplus q = q 
\oplus p$  and $p \oplus (q \oplus r) = (p \oplus q) \oplus r$\footnote{With 
one side defined  iff the other is.}; \item[(2)] $p \oplus p$ is defined only 
if $p =  0$; \item[(3)] $p \oplus 0 = 0 \oplus p = p$; \item[(4)] For every 
$p \in \Pi$, there exists a unique element --- namely, $p'$ ---satisfying $p 
\oplus  p' = 1$. \end{mlist} 
For  the test space $(S,{\frak F})$ of frames of a Hilbert space $\H$,
events  are simply orthonormal set of vectors 
in $\H$, and two events are perspective iff they have the same closed span. Hence, 
we can identify $\Pi(S,{\frak F})$ with the set of closed subspaces of $\H$, with $\oplus$ coinciding with the usual orthogonal sum operation.

{\bf 0.3 Orthoalgebras} Abstractly, a structure satisfying (1) through (4) 
above is  called an {\em orthoalgebra}. It can be shown that every 
orthoalgebra arises  canonically (though not uniquely) as $\Pi(X,{\frak A})$ 
for an algebraic test space $(X,{\frak A})$.  Indeed, if $L$ is an orthoalgebra, let 
$X_{L} = L \setminus \{0\}$ and let ${\frak A}_{L}$ denote the set of finite 
subsets $E = \{e_{1},...,e_{n}\}$ of $L \setminus {0}$ for which $e_{1} 
\oplus \cdots \oplus e_{n}$ exists and equals $1$. Then $(X_{L}, {\frak 
A}_{L})$ is an algebraic test space with logic canonically isomorphic to $L$. 

Any orthoalgebra $L$ carries a natural partial order, defined by setting $p 
\leq q$ iff there exists some $r \in L$ with $p \perp r$ and $p \oplus r =  
q$. With respect to this ordering, the mapping $p \mapsto p'$ is an  
orthocomplementation.

{\bf 0.4 Proposition [3]}: { \em If $L$ is an
orthoalgebra, the following are  equivalent:
\begin{mlist}
 \item[(a)] $L$ is {\em orthocoherent}, i.e., for all pairwise orthogonal 
elements $p, q, r \in L$, $p \oplus q \oplus r$ exists.
\item[(b)] $p \oplus q = p \vee q$ for all $p \perp q$ in $L$
\item[(c)] $(L,\leq,')$ is an orthomodular poset
\end{mlist}}
       
Note also that if $(L, \leq, ')$ is any orthoposet, the
partial binary  operation of orthogonal join --- that is, $p
\oplus q = p \vee q$ for 
$p \leq q'$ -- is associative iff $L$ is orthomodular, in
which case, 
$(L, \oplus)$ is an orthoalgebra, the natural order on which
coincides with  the given order on $L$ [11]. Thus, orthomodular 
posets and orthomodular lattices can be regarded as 
essentially the same things as orthocoherent orthoalgebras and 
lattice-ordered orthoalgebras, respectively.\\

{\large \sc 1. Background on the Vietoris Topology}

General references for this section are [7]  
and [9]. If $X$ is any topological space, let $2^X$ denote the set
of all closed subsets of $X$.  If $A \subseteq X$, let 
\[[A] := \{ F \in 2^{X} | F \cap A \not = \emptyset\}.\]
Clearly, $[A \cap B] \subseteq [A] \cap [B]$ and
$\bigcup_{i} [A_{i}] =  [\bigcup_{i} A_{i}]$. 
The {\em Vietoris topology} on $2^{X}$ is the coarsest topology
in which $[U]$ is open if $U \subseteq X$ is open and $[F]$ is closed if $F \subseteq X$ is closed. 
\footnote{In particular, $\emptyset$ is an isolated point of $2^{X}$. 
Many authors omit $\emptyset$ from $2^{X}$.} 
Thus, if $U$ is open, so is $(U) := [U^{c}]^{c} = \{ F \in 2^{X} | F 
\subseteq U\}$. Let ${\cal B}$ be any basis for the topology on $X$: then 
the collection of sets of the form 
\[\langle U_{1},...,U_{n} \rangle \ := \ [U_{1}] \cap \cdots \cap [U_{n}] \cap 
\left ( \bigcup_{i=1}^{n} U_{i} \right ) \]
with $U_{1},...,U_{n}$ in ${\cal B}$, is a basis for the Vietoris topology 
on $2^{X}$. Note that $\langle U_{1},...,U_{n} \rangle$ consists of all 
closed sets contained in $\bigcup_{i=1}^{n} U_{i}$ and meeting each set 
$U_{i}$ at least once.

If $X$ is a compact metric space, then the Vietoris topology on $2^{X}$ is 
just that induced by the Hausdorff metric. Two classical results concerning 
the Vietoris topology are {\em Vietoris' Theorem}: $2^{X}$ is compact 
iff $X$ is compact, and {\em Michael's Theorem}: a (Vietoris) compact union of 
compact sets is compact.\footnote{More precisely, if ${\cal C}$ is a compact 
subset of $2^{X}$ with each $C \in {\cal C}$ compact, then $\bigcup_{C \in {\cal C}} C$ is 
again compact.}
                                        
The operation $\cup : 2^{X} \times 2^{X} \rightarrow 2^{X}$ is also Vietoris 
continuous, since \[\cup^{-1}([U]) = \{ (A,B) | A \cup B \in [U]\} = ([U] 
\times 2^{X}) \cup (2^{X} \times [U]),\] 
which is open if $U$ is open and closed if $U$ is closed. In particular, for any fixed closed set 
$A$, the mapping $f_{A} : 2^{X} \rightarrow 2^{X}$ given by $f_{A} : B \mapsto A 
\cup B$ is continuous. Notice also that the mapping $\pi : 2^{X} \times 2^{X} 
\rightarrow 2^{X \times X}$ given by $\pi(A,B) = A \times B$ is continuous, 
as $\pi^{-1}([U \times V]) = [U] \times [V]$ and $\pi^{-1}((U \times V)) 
= (U) \times (V)$.

Henceforth, we regard any collection $\frak A$ of closed subsets of a 
topological space $X$ as a subspace of $2^{X}$. In the special case in which 
${\frak A}$ is a collection of finite sets of uniformly bounded cardinality, 
say $|E| < n$ for every $E \in {\frak A}$, there is a more direct 
approach to topologizing ${\frak A}$ that bears discussion. Let ${\frak A}^{o} 
\subseteq X^{n}$ denote the space of {\em ordered} versions 
$(x_{1},...,x_{n})$ of sets $\{x_{1},...,x_{n}\} \in {\frak A}$, with the 
relative product topology. We can give ${\frak A}$ the quotient topology 
induced by the natural surjection $\pi : {\frak A}^{o} \rightarrow {\frak A}$ 
that  ``forgets" the order. The following is doubtless well-known, but I 
include the short proof for completeness. 

{\bf 1.1 Proposition:} {\em Let $X$ be Hausdorff and ${\frak A}$, a 
collection of non-empty finite subsets of $X$ of cardinality $\leq n$ (with 
the Vietoris topology).
Then the canonical surjection $\pi : {\frak A}^{o} \rightarrow {\frak A}$ 
is an open continuous map. Hence, the Vietoris topology on ${\frak A}$ 
coincides with the quotient topology induced by $\pi$. }

Proof: Let $U_{1},...,U_{n}$ be open subsets of $X$. Then 
$\pi((U_{1} \times \cdots \times U_{n})\cap {\frak A}^{o}) = \langle 
U_{1},...,U_{n}\rangle \cap {\frak A}$, 
so $\pi$ is open. Also 
\[\pi^{-1}(\langle U_{1},...,U_{n}\rangle \cap {\frak A})  \ = \ 
\bigcup_{\sigma} (U_{\sigma(1)} \times \cdots \times U_{\sigma(n)}) \cap 
{\frak A}^{o},\]
where $\sigma$ runs over all permutations of $\{1,2,...,n\}$, so $\pi$ is 
continuous. It follows immediately that the quotient and Vietoris topologies on 
${\frak A}$ coincide. $\Box$\\

{\large \sc 2. Topological Test Spaces}

We come now to the subject of this paper. 

{\bf 2.1 Definition:} A {\em topological test space} is a
test space $(X, {\frak A})$  where $X$ is a Hausdorff 
space and the relation $\perp$ is closed in $X \times X$. 
               
{\bf 2.2 Examples}\\ (a) Let $\H$ be a Hilbert space. Let $S$ be the unit 
sphere of $\H$, in any topology making the inner product continuous. Then the 
test space $(S,{\frak F})$ defined above is a topological test space, since 
the orthogonality relation is closed in $S^{2}$.\\ (b) 
Suppose that $X$ is Hausdorff, that every $E \in {\frak A}$ is finite, and that $(X,{\frak A})$ 
supports a set $\Gamma$ of {\em continuous} probability weights that are 
$\perp$-separating in the sense that $p \not \perp q$ iff $\exists \omega \in 
\Gamma$ with $\omega(p) + \omega(q) > 1$. Then $\perp$ is closed in $X^{2}$, 
so again $(X,{\frak A})$ is a topological test space.\\ 
(c) Let $L$ be any 
topological orthomodular lattice [1]. The mapping $\phi: L^{2} \rightarrow 
L^{2}$ given by $\phi(p,q) = (p,p \wedge q')$ is continuous, and $\perp = 
\phi^{-1}(\Delta)$ where $\Delta$ is the diagonal of $L^{2}$. Since $L$ is 
Hausdorff, $\Delta$ is closed, whence, so is $\perp$. Hence, the test space 
$(L\setminus \{0\},{\frak A}_{L})$ (as described in 0.3 above) is 
topological.

The following Lemma collects some basic facts about topological test spaces 
that will be used freely in the sequel. 

{\bf 2.3 Lemma:} {\em Let $(X, {\frak A})$ be a topological
test space. Then 
\begin{mlist}
\item[(a)] Each point $x \in X$ has an open neighborhood containing no 
two orthogonal outcomes. (We shall call such a neighborhood {\em totally non-
orthogonal}.)
\item[(b)] For every set $A \subseteq X$, 
$A^{\perp}$ is closed.
\item[(c)] Each pairwise orthogonal subset of $X$ is discrete 
\item[(d)] Each pairwise orthogonal subset of $X$ is closed. 
\end{mlist}}

Proof:  (a) Let $x \in X$. Since $(x,x) \not \in \perp$ and $\perp$ is 
closed, we can find open sets $V$ and $W$ about $x$ with 
$(V \times W) \cap \perp = \emptyset$. Taking $U = V \cap W$ gives 
the advertised result. 

(b) Let $y \in X \setminus x^{\perp}$. Then $(x,y)
\not \in \perp$. Since  the latter is closed, there exist
open sets $U, V \subseteq X$ with $(x,y) 
\in U \times V$ and $(U \times V) \cap \perp = \emptyset$. 
Thus,  no element of $V$ lies orthogonal to any element of
$U$; in particular, we  have $y \in V \subseteq X \setminus
x^{\perp}$. Thus, $X \setminus x^{\perp}$  is open, i.e.,
$x^{\perp}$ is closed. It now follows that for any  set 
$A \subseteq X$, the set $A^{\perp} = \bigcap_{x \in A}
x^{\perp}$ is closed.

(c) Let $D$ be pairwise orthogonal. Let $x \in D$: by part
(b), $X \setminus x^{\perp}$ is  open, whence, $\{x\} = D
\cap (X \setminus x^{\perp})$ is relatively open  in
$D$. Thus, $D$ is discrete. 

(d) Now suppose $D$ is pairwise orthogonal, and let $z \in \overline{D}$: if 
$z \not \in D$, then for every open neighborhood $U$ of $z$, $U \cap D$ is 
infinite; hence, we can find distinct elements $x, y \in D \cap U$. Since $D$ 
is pairwise orthogonal, this tells us that $(U \times U) \cap \perp \not = 
\emptyset$. But then $(x,x)$ is a limit point of $\perp$. Since 
$\perp$ is closed, $(x,x) \in \perp$, which is a contradiction. Thus, 
$z \in D$, i.e., $D$ is closed. $\Box$
                                      
It follows in particular that every test $E \in {\frak A}$ and every 
event $A \in {\cal E}(X,{\frak A})$ is a closed, discrete subset of $X$. 
Hence, we may construe $\frak A$ and ${\cal E}(X,{\frak A})$ of 
as subspaces of $2^{X}$ in the Vietoris topology. 

A test space $(X,{\frak A})$ is {\em  locally finite} iff each test $E \in 
{\frak A}$ is a finite set. We shall say that a test space $(X,{\frak A})$ is of {\em rank 
$n$} if $n$ is the maximum cardinality of a test in ${\frak A}$. If all tests 
have cardinality {\em equal} to $n$, then $(X,{\frak A})$ is {\em $n$-uniform}. 
                                                                
{\bf 2.4 Theorem:} {\em Let $(X,{\frak A})$ be a topological test space 
with $X$ compact. Then all pairwise orthogonal subsets of $X$ are finite, 
and of uniformly bounded size. In particular, $\frak A$ is of finite rank.}

Proof: By Part (a) of Lemma 2.3, every point $x \in X$ is contained in some totally non-
orthogonal open set. Since $X$ is compact, a finite number of these, say 
$U_{1},...,U_{n}$, cover $X$. A pairwise orthogonal set $D \subseteq X$ 
can meet each $U_{i}$ at most once; hence, $|D| \leq n$. $\Box$.

For locally finite topological test spaces, the Vietoris topology on the 
space of events has a particularly nice description. Suppose $A$ is a finite 
event: By Part (a) of Lemma 2.3, we can find for each $x \in A$ a totally 
non-orthogonal open neighborhood $U_{x}$. Since $X$ is 
Hausdorff and $A$ is finite, we can arrange for these to be disjoint from one 
another. Consider now the Vietoris-open neighborhood ${\cal V} = \langle U_{x}, x \in 
A \rangle \cap {\cal E}$ of $A$ in ${\cal E}$: an event $B$  belonging to 
$\cal V$ is contained in $\bigcup_{x \in A} U_{x}$ and meets each $U_{x}$ in 
at least one point; however, being pairwise orthogonal, $B$ can meet each 
$U_{x}$ {\em at most} once.  Thus, $B$ selects {\em exactly one} point from 
each of the disjoint sets $U_{x}$ (and hence, in particular, $|B| = |A|$). 
Note that, since the totally non-orthogonal sets form a basis for 
the topology on $X$, open sets of the form just described form a basis for 
the Vietoris topology on ${\cal E}$. 
                           
As an immediate consequence of these remarks, we have the following:
  
{\bf 2.5 Proposition:} {\em Let $(X,{\frak A})$ be locally  finite. Then the 
set ${\cal E}_{n}$ of all events of a given cardinality $n$ is clopen in 
${\cal  E}(X,{\frak A})$.}
                                       
A test space $(X,{\frak A})$ is {\em UDF} ({\em unital, 
dispersion-free}) iff for ever $x \in X$ there exists a $\{0,1\}$-valued 
state $\omega$ on $(X,{\frak A})$ with $\omega(x) = 1$. Let $U_{1},...,U_{n}$ be pairwise 
disjoint totally non-orthogonal open sets, and and let 
${\cal U} = \langle U_{1},...,U_{n} \rangle$: then ${\cal U}$ can be 
regarded as a UDF test space (each $U_{i}$ selecting one outcome from each 
test in ${\cal V}$). The foregoing considerations thus have the further interesting consequence that any locally finite topological test 
space is {\em locally UDF}. In particular, for such test spaces, the 
existence or non-existence of dispersion-free states will depend entirely on 
the {\em global} topological structure of the space. 
                                                          
If $(X,{\frak A})$ is a topological test space, let $\overline{\frak A}$ 
denote the (Vietoris) closure of ${\frak A}$ in $2^{X}$. We are going to 
show that $(X,\overline{{\frak A}})$ is again a topological test space, having 
in fact the same orthogonality relation as $(X,{\frak A})$. If $(X,{\frak 
A})$ is of finite rank, moreover, $(X,\overline{\frak A})$ has the same 
states as $(X,{\frak A})$. 
                           
{\bf 2.6 Lemma:} {\em Let $(X,{\frak A})$ be any topological test 
space, and let $E \in \overline{\frak A}$. Then $E$ is pairwise orthogonal 
(with respect to the orthogonality induced by $\frak A$).}  
        
Proof: Let $x$ and $y$ be two distinct points of $E$. Let $U$ and $V$ be 
disjoint neighborhoods of $x$ and $y$ respectively, and let 
$(E_{\lambda})_{\lambda \in \Lambda}$ be a net of closed sets in ${\frak A}$ 
converging to $E$ in the Vietoris topology. Since $E \in [U] \cap [V]$,  we 
can find $\lambda_{U,V} \in \Lambda$ such that $E_{\lambda} \in [U] \cap [V]$ 
for all $\lambda \geq \lambda_{U,V}$. In particular, we can find 
$x_{\lambda_{U,V}} \in E_{\lambda_{U,V}} \cap U$ and $y_{\lambda_{U,V}}\in 
E_{\lambda_{U,V}} \cap V$. Since $U$ and $V$ are disjoint, 
$x_{\lambda_{U,V}}$ and $y_{\lambda_{U,V}}$ are distinct, and hence, -- since 
they belong to a common test $E_{\lambda}$ -- orthogonal. This gives us a net 
$(x_{\lambda_{U,V}},y_{\lambda_{U,V}})$ in $X \times X$ converging to $(x,y)$ 
and with $(x_{\lambda_{U,V}},y_{\lambda_{U,V}}) \in \perp$. Since $\perp$ is 
closed, $(x,y) \in \perp$, i.e., $x \perp y$. $\Box$ 

It follows that the orthogonality relation on $X$ induced by $\overline{\frak 
A}$ is the same as that induced by ${\frak A}$. In particular, 
$(X,\overline{\frak A})$ is again a topological test space. 

Let ${\cal F}_{n}$ denote the set of finite subsets of $X$ having $n$ or 
fewer elements. 
        
{\bf 2.7 Lemma:} {\em Let $X$ be Hausdorff. Then for every $n$, 
\begin{mlist}
\item[(a)] ${\cal 
F}_{n}$ is closed in $2^{X}$.
\item[(b)] If $f : X \rightarrow {\Bbb R}$ is continuous, then so 
is the mapping $\hat{f} : {\frak F}_{n} \rightarrow {\Bbb R}$ given 
by $\hat{f}(A) := \sum_{x \in A} f(x)$.\end{mlist}}

Proof: (a) Let $F$ be a closed set (finite or infinite) of cardinality 
greater than $n$. Let $x_{1},...,x_{n+1}$ be distinct elements of $F$, and 
let $U_{1},....,U_{n}$ be pairwise disjoint open sets with $x_{i} \in U_{i}$ 
for each $i = 1,...,n$. Then no closed set in ${\cal U} := [U_{1}] \cap 
\cdots \cap [U_{n}]$ has fewer than $n+1$ points -- i.e, ${\cal U}$ is an 
open neighborhood of $F$ disjoint from ${\cal F}_{n}$. This shows that $2^{X} 
\setminus {\cal F}_{n}$ is open, i.e., ${\cal F}_{n}$ is closed. 
                                                   
(b) By proposition 1.1, ${\frak F}_{n}$ is the quotient space of 
$X^{n}$ induced by the surjection  
surjection $q: (x_{1},...,x_{n}) \mapsto \{x_{1},...,x_{n}\}$.  The mapping $\overline{f} : X^{n} \rightarrow {\Bbb R}$ 
given by $(x_{1},...,x_{n}) \mapsto \sum_{i=1}^{n} f(x_{i})$ is plainly 
continuous; hence, so is $\hat{f}$. $\Box$ 

{\bf 2.8 Proposition:} {\em Let $(X,{\frak A})$ be a rank-$n$ (respectively, 
$n$-uniform) test space. Then 
$(X,\overline{\frak A})$ is also a rank-$n$ (respectively, $n$-uniform) test 
space having the same continuous states as $(X,{\frak A})$. }

Proof: If ${\frak A}$ is rank-$n$, then ${\frak A} \subseteq {\frak F}_{n}$. 
Since the latter is closed, $\overline{\frak A} \subseteq {\frak F}_{n}$ 
also.  Note that if ${\frak A}$ is $n$-uniform and $E \in \overline{\frak 
A}$, then any net $E_{\lambda} \rightarrow E$ is eventually in bijective 
correspondence with $E$, by Proposition 2.5. Hence, $(X,\overline{\frak A})$ is 
also $n$-uniform. Finally, every continuous state on $(X,{\frak A})$ lifts to 
a continuous state on $(X,\overline{\frak A})$ by Lemma 2.7 (b). $\Box$\\

{\large \sc 3.  The Logic of a Topological Test Space}

In this section, we consider the logic $\Pi = \Pi(X,{\frak A})$ of 
an algebraic test space $(X,{\frak A})$. We endow this  
with the quotient topology induced by the canonical surjection 
$p : {\cal E} \rightarrow \Pi$ (where ${\cal E} = {\cal E}(X,{\frak A})$ 
has, as usual, its Vietoris topology). Our aim is to find conditions on $(X,{\frak A})$ that will 
guarantee reasonable continuity properties for the 
orthogonal sum operation and the orthocomplement. In this connection, 
we advance the following

{\bf 3.1 Definition:}  A {\sl topological orthoalgebra} is an orthoalgbra 
$(L,\perp,\oplus,0,1)$ in which $L$ is a topological space, the 
relation $\perp \subseteq L^{2}$ is closed, and the mappings 
$\oplus : \perp \rightarrow L$ and $' : L \rightarrow L$ are continuous.

A detailed study of topological orthoalgebras must wait for another 
paper. However, it is worth mentioning here that, while every topological orthomodular lattice 
is a topological orthoalgebra, there exist lattice-ordered topological 
orthoalgebras in which the meet and join are discontinuous -- e.g., the orthoalgebra $L(\H)$ of closed 
subspaces of a Hilbert space, in its operator-norm topology.

{\bf 3.2 Lemma:} {\em Let $(L,\perp,\oplus,0,1)$ be a topological 
orthoalgebra. Then 
\begin{mlist}
\item[(a)] The order relation $\leq$ is closed in $L^{2}$ 
\item[(b)] $L$ is a Hausdorff space.\end{mlist}}

Proof: For (a), note that $a \leq b$ iff $a \perp b'$. Thus, $\leq \ = \ f^{-
1}(\perp)$ where $f : L \times L \rightarrow L \times L$ is the continuous 
mapping $f(a,b) = (a,b')$. Since $\perp$ is closed, so is $\leq$. The second 
statement now follows by standard arguments (cf. Nachbin [10]). $\Box$

We now return to the question: when is the logic of a topological test space, 
in the quotient topology, a topological orthoalgebra? 

{\bf 3.3 Lemma:} {\em Suppose 
${\cal E}$ is closed in $2^{X}$. Then 
\begin{mlist}
\item[(a)]
The orthogonality relation $\perp_{\cal E}$ on ${\cal E}$ is closed in ${\cal E}^{2}$.
\item[(b)] 
The mapping $\cup : \perp_{\cal E} \rightarrow {\cal E}$ is continuous
\end{mlist}}
                              
Proof: 
The mapping ${\cal E}^2 \rightarrow 2^{X}$ given by 
$(A,B) \mapsto A \cup B$ is continuous; hence, if ${\cal E}$ is closed in 
$2^{X}$, then so is the set $\C := 
\{(A,B) \in {\cal E}^{2} | A \cup B \in {\cal E}\}$ of {\em compatible} pairs 
of events. It will suffice to show that the set $\O := \{ (A,B) \in {\cal E} | A \subseteq 
B^{\perp}\}$ is also closed, since $\perp = \C \cap \O$. But 
$(A,B) \in \O$ iff $A \times B \subseteq \perp$, i.e., 
$\O = \pi^{-1}((\perp)) \cap {\cal E}$ where $\pi : 2^{X} \times 2^{X} 
\rightarrow 2^{X \times X}$ is the product mapping $(A,B) \mapsto A \times 
B$. As observed in section 1, this mapping is continuous, and since 
$\perp$ is closed in $2^{X \times X}$, so is $(\perp)$ in $2^{X \times X}$. 
Statement (b) follows immediately from the Vietoris continuity of $\cup$. 
$\Box$

{\em Remarks:} The hypothesis that ${\cal E}$ be closed in $2^{X}$ is not used 
in showing that the relation $\O$ is closed. If $(X,{\frak A})$ is {\em 
coherent} [10], then $\O = \perp$, so in this case, the hypothesis can be avoided 
altogether. On the other hand, if $X$ is compact and ${\frak A}$ is closed, 
then ${\cal E}$ will also be compact and hence, closed. (To see this, note 
that if $X$ is compact then by Vietoris' Theorem, $2^{X}$ is compact. Hence, 
so is the closed set $(E) = \{ A \in 2^{X} | A \subseteq E\}$ for each $E \in 
{\frak A}$.  The mapping $2^{X} \rightarrow 2^{2^{X}}$ given by $E \mapsto (E)$ is easily seen to 
be continuous. Since $\frak A$ is closed, hence compact,  
in  $2^{X}$, it follows that $\{(E) | E \in {\frak A}\}$ is a compact subset of $2^{2^{X}}$. 
By Michael's theorem, ${\cal E} = \bigcup_{E \in {\frak A}} (E)$ is compact, hence closed, 
in $2^X$.) 
                                                                 
In order to apply Lemma 3.3 to show that $\perp \subseteq \Pi^{2}$ is closed 
and $\oplus : \perp \rightarrow \Pi$ is continuous, we would like to have the 
canonical surjection $p : {\cal E} \rightarrow \Pi$ open. The following 
condition is sufficient to secure this, plus  the continuity of the 
orthocomplementation $' : \Pi \rightarrow \Pi$. 

{\bf 3.3 Definition:} Call a topological test space $(X,{\frak A})$ is {\em stably 
complemented} iff for any open set ${\cal U}$ in ${\cal E}$, the set 
${\cal U}^{\oc}$ of events complementary to events in ${\cal U}$ is again 
open.

{\em Remark:} If $\H$ is a finite-dimensional Hilbert space, it can be shown that the 
corresponding test space $(S,{\frak F})$ of frames is 
stably complemented [12].

{\bf 3.5 Lemma:} {\em Let $(X, {\frak A})$ be a topological test space, 
and let $p : {\cal E} \rightarrow \Pi$ be the canonical quotient mapping 
(with $\Pi$ having the quotient topology). Then the following are 
equivalent: 
\begin{mlist}
\item[(a)] $(X,{\frak A})$ is stably complemented 
\item[(b)] The mapping $p: {\cal E} \rightarrow \Pi$ is open and the mapping $' : \Pi \rightarrow \Pi$ 
is continuous.\end{mlist}}

Proof: Suppose first that $(X,{\frak A})$ is stably complemented, 
and let ${\cal U}$ be an open set in ${\cal E}$. Then 
\begin{eqnarray*}
p^{-1}(p({\cal U}) ) & = & \{ A \in {\cal E} | \exists B \in {\cal U} A \sim 
B\}\\
& = & \{ A \in {\cal E} | \exists C \in {\cal U}^{\oc} A \oc C\}\\
& = & \left ( {\cal U}^{\oc} \right )^{\oc}\end{eqnarray*}
which is open. Thus, $p({\cal U})$ is open. Now note that $' : \Pi \rightarrow \Pi$ is 
continuous iff, for every open set $V \subseteq \Pi$, the set $V' = \{p' | p \in V\}$ is 
also open. But $p^{-1}(V') = (p^{-1}(V))^{\oc}$: since $p$ is continuous and 
$(X,{\frak A})$ is stably complemented, this last is open. Hence, $V'$ is open. 

For the converse, note first that if $'$ is continuous, it is also open 
(since $a'' = a$ for all $a \in \Pi$). Now 
for any open set ${\cal U} \subseteq {\cal E}$, ${\cal U}^{\oc} = 
p^{-1}(p({\cal U})')$: Since $p$ and $'$ are continuous open mappings, 
this last is open as well. $\Box$ 
                    
{\bf 3.6 Proposition:} {\em Let $(X,{\frak A})$ be a stably complemented 
algebraic test space with ${\cal E}$ closed. Then $\Pi$ is a topological 
orthoalgebra.}
                                              
Proof: Continuity of $'$ has already been established. 
We show first that $\perp \subseteq \Pi^{2}$ is closed. If 
$(a,b) \not \in \perp$, then for all $A \in p^{-1}(a)$ and $B \in p^{-1}(b)$, 
$(A,B) \not \in \perp_{\cal E}$. The latter is closed, by Lemma 3.3 (a); hence, 
we can find Vietoris-open neighborhoods ${\cal U}$ and ${\cal V}$ of $A$ and $B$, respectively, 
with $({\cal U} \times {\cal V}) \cap \perp_{\cal E} = \emptyset$. 
Since $p$ is open, $U := p({\cal U})$ and $V := p({\cal V})$ are open 
neighborhoods of $a$ and $b$ with $(U \times V) \cap \perp = \emptyset$. 
To establish the continuity of $\oplus : \perp \rightarrow \Pi$, 
let $a \oplus b = c$ and let $A \in p^{-1}(a), B \in p^{-1}(B)$ 
and $C \in p^{-1}(c)$ be representative events. Note that $A \perp B$ 
and $A \cup B = C$. Let $W$ be an open set containing $c$: then 
${\cal W} := p^{-1}(W)$ is an open set containing $C$. By Lemma 3.3 (b), $\cup : \perp_{\cal 
E} \rightarrow {\cal E}$ is continuous; hence, we can find open 
sets ${\cal U}$ about $A$ and ${\cal V}$ about $B$ with $A_{1} \cup B_{1} \in {\cal
W}$ for every $(A_{1},B_{1}) \in ({\cal U} \times {\cal V}) \cap \perp_{\cal E}$.
Now let $U = p({\cal U})$ and $V = p({\cal V})$: these are open neighborhoods 
of $a$ and $b$, and for every $a_{1} \in U$ and $b_{1} \in V$ with $a_{1} \perp b_{1}$, 
$a_{1} \oplus b_{1}  \in p(p^{-1}(W)) = W$ (recalling here that $p$ is surjective).
Thus, $(U \times V) \cap \perp \subseteq \oplus^{-1}(W)$, so $\oplus$ is 
continuous. $\Box$\\
                                               
{\large \sc 5.  Semi-classical Test Spaces}

From a purely combinatorial point of view, the simplest test spaces are those  
in which distinct tests do not overlap. Such test spaces are said to be {\em  
semi-classical}. In such a test space, the relation of perspectivity is the 
identity relation on events; consequently, the logic of a semi-classical test  
space $(X,{\frak A})$ is simply the horizontal sum of the boolean algebras 
$2^{E}$, $E$  ranging over ${\frak A}$. A state on a semi-classical test 
space $(X,{\frak  A})$ is simply an assignment to each $E \in {\frak A}$ of a 
probability  weight on $E$. (In particular, there is no obstruction to constructing 
``hidden variables" models for states on such test spaces.) 

Recent work of D. Meyer [8] and of R. Clifton and A. Kent [2] has shown that the
test space $(S(\H),{\frak F}(\H))$ associated with a finite-dimensional Hilbert space 
contains (in  our language) a dense semi-classical sub-test space. To conclude this paper, I'll show that 
the this result in fact holds for a large and rather natural class of topological test spaces.  
            
{\bf 4.1 Lemma:} {\em Let $X$ be any Hausdorff (indeed, $T_{1}$) space, and let $U \subseteq X$ be a 
dense open set. Then $(U) = \{ F \in 2^{X} | F \subseteq U\}$ is a dense open set in $2^{X}$.}

Proof: Since sets of the form $\langle U_{1},....,U_{n}\rangle$, 
$U_{1},...,U_{n}$ open in $X$, form a basis for the Vietoris topology 
on $2^{X}$, it will suffice to show that $(U) \cap \langle 
U_{1},...,U_{n} \rangle \not = 0$ for all choices of non-empty opens 
$U_{1},...,U_{n}$. Since $U$ is dense, we can select for each $i = 1,...,n$ a 
point $x_{i} \in U \cap U_{i}$. The finite set $F := \{x_{1},...,x_{n}\}$ is 
closed (since $X$ is $T_{1}$), and by construction lies in $(U) \cap \langle 
U_{1},...,U_{n} \rangle$. $\Box$ 

{\bf 4.2 Corollary:} {\em Let $(X,{\frak A})$ be any topological test space 
with $X$ having no isolated points, and let $E$ be any test in ${\frak A}$. 
Then open set $(E^{c}) = [E]^{c}$ of tests disjoint from $E$ is dense in 
${\frak A}$.} 

Proof: Since $E$ is a closed set, its complement $E^{c}$ is an open set; 
since $E$ is discrete and includes no isolated point, $E^{c}$ is dense. The 
result follows from the preceding lemma. $\Box$ 

{\bf 4.3 Theorem:} {\em Let $(X,{\frak A})$ be a topological test space with 
$X$ (and hence, ${\frak A}$) second countable, and without isolated points. 
Then there  exists a countable, pairwise-disjoint sequence $E_{n} \in {\frak 
A}$ such  that (i) $\{E_{n}\}$ is dense in ${\frak A}$, and (ii) $\bigcup_{n} 
E_{n}$ is  dense in $X$.} 

\newpage
Proof: Since it is second countable, ${\frak A}$ has a
countable basis of open sets ${\cal W}_{k}$, $k \in {\Bbb N}$. Selecting an element $F_{k} \in {\cal W}_{k}$ for 
each $k \in {\Bbb N}$, we obtain a countable dense subset of
${\frak A}$. We shall construct a countable dense pairwise-disjoint 
subsequence $\{E_{j}\}$ of $\{F_{k}\}$. Let 
$E_{1} = F_{1}$.  By Corollary 4.2, $[E_{1}]^{c}$ is a dense open set; hence, it has 
a  non-empty intersection with ${\cal W}_{2}$. As $\{F_{k}\}$ is
dense, there exists an  
index $k(2)$ with $E_{2} := F_{k(2)} \in {\cal W}_{2} \cap [E_{1}]^{c}$.
We now have $E_{1} \in W_{1}$, $E_{2} \in {\cal W}_{2}$, and 
$E_{1} \cap E_{2}  = \emptyset$. Now proceed
recursively: Since 
$[E_{1}]^{c} \cap [E_{2}]^{c} \cap \cdots \cap
[E_{j}]^{c}$ is a  dense open and ${\cal W}_{j+1}$ is a
non-empty open, they have a non-empty  intersection; hence,
we can select $E_{j+1} = F_{k(j+1)}$ belonging to this  intersection.
This will give us a test belonging to ${\cal W}_{j+1}$ but
disjoint  from each of the pairwise disjoint sets
$E_{1},...,E_{j}$. Thus, we obtain a sequence
$E_{j} := F_{k(j)}$ of pairwise disjoint tests,  one of
which lies in each non-empty basic open set ${\cal W}_{j}$ -- and which
are, therefore,  dense. 

For the second assertion, it now suffices to notice that for
each open set 
$U \subseteq X$, $[U]$ is a non-empty open in ${\frak A}$, and hence
contains some 
$E_{j}$. But then $E_{j} \cap U \not = \emptyset$, whence,
$\bigcup_{j}  E_{j}$ is dense in $X$. $\Box$\\

{\large \sc References} 

[1] Choe, T. H., Greechie, R. J., and Chae, Y., {\em Representations of locally compact 
orthomodular lattices}, Topology and its Applications {\bf 56} (1994) 165-173
 
[2] Clifton, R., and Kent, {\em Simulating quantum mechanics with 
non-contextual hidden variables}, The Proceedings of the Royal Society of 
London A {\bf 456} (2000) 2101-2114 

[3] Foulis, D. J., Greechie, R. J., 
and Ruttimann, G. T., {\em Filters and supports on  orthoalgebras} International 
Journal of Theoretical Physics {\bf 31} (1992) 789-807

[4] Foulis, D. J., Greechie, R. J., and 
Ruttimann, G. T., {\em Logico-algebraic structures  II: supports 
on test spaces}, International Journal of Theoretical Physics  
{\bf 32} (1993) 1675-1690

[5] Foulis, D. J., and Randall, C.H., {\em A mathematical language for 
quantum physics}, in H. Gruber et al. (eds.), {\em Les fondements de la 
m\'{e}chanique quantique}, AVCP: Lausanne (1983)

[6] Gleason, A., {\em Measures on the closed subspaces of Hilbert 
space}, Journal of Mathematics and Mechanics {\bf 6} (1957) 885-894 

[7] Illanes, A., and Nadler, S. B., {\bf Hyperspaces}, Dekker: New York 
(1999)

[8] Meyer, D., {\em Finite precision measurement nullifies the Kochen-Specker theorem}, Physical Review Letters {\bf 83} (1999) 3751-3754

[9] Michael, E., {\em Topologies on spaces of subsets}, 
Transactions of the American Mathematical Society {\bf 7} (1951) 152-182

[10] Nachbin, L., {\bf Topology and Order}, van Nostrand: Princeton 1965

[11] Wilce, A., {\em Test Spaces and Orthoalgebras}, in Coecke et al (eds.)
{\bf Current Research in Operational  Quantum Logic}, Kluwer: Dordrecht 
(2000)

[12] Wilce, A., {\em Symmetry and Compactness in Quantum Logic}, in preparation.

\end{document}